\documentclass{article}

\usepackage{amsfonts}
\usepackage{amsmath}
\usepackage{epsfig}

\def\beq{\begin{equation}}
\def\eeq{\end{equation}}
\def\bea{\begin{eqnarray}}
\def\eea{\end{eqnarray}}
\def\bq{\begin{quote}}
\def\eq{\end{quote}}

\newcommand{\gsim}{\lower.7ex\hbox{$\;\stackrel{\textstyle>}{\sim}\;$}}
\newcommand{\lsim}{\lower.7ex\hbox{$\;\stackrel{\textstyle<}{\sim}\;$}}

\begin{document}

\title{\vspace*{-2cm}\hfill\hspace{4.1in}{\small KIAS--P07031, MCTP--07--18} \\
\vspace*{1cm}
Quintessential Kination and Leptogenesis}
\author{\vspace*{0.5cm} Eung Jin Chun$^{1,2}$ and Stefano Scopel$^1$\\
$^1$Korea Institute for Advanced Study\\
207-43 Cheongryangri-dong Dongdaemun-gu\\
Seoul 130-722, Korea \\
$^2$Michigan Center for Theoretical Physics\\
Department of Physics, University of Michigan\\
Ann Arbor, MI 48109, USA }
\date{}

\maketitle

\begin{abstract}
Thermal leptogenesis induced by the CP-violating decay of a
right-handed neutrino (RHN) is discussed in the background of
quintessential kination, i.e., in a cosmological model where the
energy density of the early Universe is assumed to be dominated by the
kinetic term of a quintessence field during some epoch of its
evolution. This assumption may lead to very different observational
consequences compared to the case of a standard cosmology where the
energy density of the Universe is dominated by radiation. We show
that, depending on the choice of the temperature $T_r$ above which
kination dominates over radiation, any situation between the strong
and the super--weak wash--out regime are equally viable for
leptogenesis, even with the RHN Yukawa coupling fixed to provide the
observed atmospheric neutrino mass scale $\sim 0.05$ eV. For $M\lsim
T_r \lsim M/100$, i.e., when kination stops to dominate at a time
which is not much later than when leptogenesis takes place, the
efficiency of the process, defined as the ratio between the produced
lepton asymmetry and the amount of CP violation in the RHN decay, can
be larger than in the standard scenario of radiation domination. This
possibility is limited to the case when the neutrino mass scale is
larger than about 0.01 eV.  The super--weak wash--out regime is
obtained for $T_r << M/100$, and includes the case when $T_r$ is close
to the nucleosynthesis temperature $\sim 1$ MeV.  Irrespective of
$T_r$, we always find a sufficient window above the electroweak
temperature $T\sim 100$ GeV for the sphaleron transition to
thermalize, so that the lepton asymmetry can always be converted to
the observed baryon asymmetry.

\end{abstract}

\section{Introduction}

Present observations strongly favour the existence of dark energy,
which contributes a fraction $\Omega_{\rm DE}\simeq$ 0.7 to the
closure density.  This dark energy component in the present universe
can be explained by modifying the standard cosmology with the
introduction of a slowly evolving scalar field called quintessence
\cite{caldwell}.  This approach has been shown to have ``tracking
solutions'' \cite{tracking} that solve the so-called coincidence
problem suffered by a pure cosmological constant, namely explaining in
a natural way why radiation and dark energy provide comparable
contributions to the energy budget of the present Universe, in spite
of having very different time evolutions. An open possibility in this
scenario is the existence of an early era of kination domination,
during which the Universe is dominated by the kinetic energy of the
quintessence field.  During this era, the expansion rate of the
Universe is larger compared to the usual radiation domination case. An
interesting consequence of this fact is that the relic abundance of a
Cold Dark Matter candidate can be significantly enhanced compared to
the canonical prediction \cite{salati}, because its decoupling time
from the plasma is anticipated. In this way, for models with high
detection rates, the relic density that is usually small in the
standard case can be brought back to the range compatible to
observation. This leads to interesting phenomenological implications
for the LHC and other future collider or astrophysics
experiments. Another interesting prediction of the kination domination
scenario in inflationary models is the absence of a measurable tensor
perturbation induced B-mode CMB polarization, which can be tested in
the next--generation CMB experiments \cite{chung}.

In this paper, we investigate the impact of the quintessential
kination scenario on the properties of the thermal leptogenesis
induced by the CP-violating decay of a right-handed neutrino (RHN),
$N$ \cite{yanagida}.  In particular, when the RHN Yukawa coupling is
fixed in order to explain the observed atmospheric neutrino mass scale
$\simeq$ 0.05 eV through the see--saw mechanism, leptogenesis is known
to proceed in the strong wash--out regime for a standard Cosmology.
As will be shown in the following, similarly to the case of the relic
density of a thermal relic, the faster expansion rate in the early
Universe due to kination dominance can modify the predictions for the
standard leptogenesis scenario, even by several orders of magnitude.
As a consequence of this, any situation between the strong and the
super--weak wash--out regime are equally viable for leptogenesis in
presence of kination domination.

The paper is organized as follows. In Section \ref{sec:kination}
the main features of the quintessential cosmological scenario are
summarized, and the basic requirements for kination domination at
early times are discussed. In Section \ref{sec:themodel} thermal
leptogenesis is discussed in the context of an MSSM model
supplemented by heavy right--handed neutrino supermultiplets with
CP--violating decays, and the relevant Boltzmann equations are
introduced. Section \ref{sec:discussion} is devoted to our
discussion, where the solutions for the Boltzmann equations of
leptogenesis for a kination dominated Universe are compared to
those for the standard radiation dominated case. We summarize our
conclusions in Section \ref{sec:conclusions}.

\section{Kination and leptogenesis}

\label{sec:kination}

Let us begin with a brief summary of the cosmological behavior of
kination. The kination regime is attained when, in the energy--momentum
tensor of the quintessence field $\phi$ (assumed here as spatially
constant): $
T_{\mu\nu}=\partial_{\mu}\phi \frac{\partial {\cal L}}{\partial
  \partial^{\nu}\phi}-g_{\mu\nu}{\cal L}$,
the kinetic term $\dot{\phi}^2/2$ dominates over the
  potential term $V(\phi)$, so that
\beq w\equiv
\frac{p}{\rho}=\frac{\frac{\dot{\phi}^2}{2}-V(\phi)}{\frac{\dot{\phi}^2}{2}+V(\phi)}\rightarrow
1. \label{eq:w} \eeq
Eq.~(\ref{eq:w}) must be compared to the
corresponding values for radiation ($w=1/3$), vacuum ($w=-1$) and
pressure--less dust ($w=0$).  The energy density of the Universe
scales like
$\rho\propto a^{-3(1+w)}$, which, in particular, implies

\bea
\rho_{rad}&\propto& a^{-4}\;\;\;\;{\rm (radiation)}\nonumber\\
\rho_{kin}&\propto& a^{-6}\;\;\;\;{\rm (kination)}.\label{eq:a_dep}
\eea

\noindent In the following, we will assume that, in the epoch after reheating
which is relevant to thermal leptogenesis, the energy
density of the Universe is dominated by the sum of the above two
components, $\rho=\rho_{rad}+\rho_{kin}$, with the
boundary condition:
\beq \rho_{kin}(T_r)=\rho_{kin}(T_r),
\label{eq:boundary}
\eeq
\noindent where the kination--radiation equality temperature $T_r$ is
in principle a free parameter, with the only constraint: $T_r \gsim$ 1
MeV, in order not to spoil big-bang nucleosynthesis.
Eqs.~(\ref{eq:a_dep},\ref{eq:boundary}) imply

\beq
\rho(T)=\rho_{rad}(T)+\rho_{rad}(T_r)\left( \frac{a_r}{a}\right )^6,
\eeq

\noindent where
$\rho_{rad}(T)=\pi^2/30 g_* T^4$. Assuming an isoentropic expansion of
the Universe, $(a^3 s)$=constant, with $s=2\pi^2/45 g_* T^3$, we get

\beq
g_* a^3 T^3=g_{*r} a_r^3 T_r^3\rightarrow \left( \frac{a_r}{a}\right
)^6=\left ( \frac{g_*}{g_{*r}}\right )^2\left (\frac{T}{T_r}\right )^6,
\eeq

\noindent where $g_*$ is the number of relativistic degrees of
freedom, while $a_r$ and $g_{*r}$ are the values of $a$ and $g_*$ at
temperature $T_r$. Thus one finds the dependence of $\rho$ on the
temperature $T$ as \beq \rho(T) =\frac{\pi^2}{30}g_* T^4 \left
(1+\frac{g_*}{g_{*r}}\left (\frac{T}{T_r}\right )^2\right
)\label{eq:rho_tot}.  \eeq The above form of the energy density drives
the expansion of the Universe through the Hubble parameter given by

\beq
H(T)=1.66 \sqrt{g_*}\frac{T^2}{m_{pl}}
\sqrt{1+\frac{g_*}{g_{*r}}\left ( \frac{T}{T_r}\right )^2},
\label{eq:H}
\eeq
\noindent where $m_{pl}=1.22 \times 10^{19}$ GeV.
By introducing the non--dimensional quantity $z\equiv M/T$, where $M$
is the mass of the heavy neutrino producing leptogenesis, the scaling
of $H$, which will be relevant to the solution of the Boltzmann
equation discussed in the next section, can be expressed in the simple
form:

\beq
H(z)=\sqrt{\frac{z^2+z_r^2}{1+z_r^2}}\frac{H_1}{z^3},
\label{eq:H_zr}
\eeq

\noindent where $H_1\equiv H(z=1)$ and  $z_r\equiv
\sqrt{\frac{g_*}{g_{*r}}}M/T_r$.

The main consequence of Eqs.~(\ref{eq:H},\ref{eq:H_zr}) is that, whenever $T>T_r$,
the expansion rate is dominated by kination, and can be much faster
compared to radiation. In particular, taking $T_r = 1$ MeV,
$g_{*r}=10.75$ and $g_*(T)=228.75$ in the supersymmetric standard
model, we get
\begin{equation}
 H(T) \simeq 0.95 \times10^4 \mbox{ GeV}
 \left(\frac{ 3.28 \mbox{ MeV}} {\sqrt{g_{*r}}\, T_r} \right)
 \left(\frac{ T}{10^6 \mbox{ GeV}} \right)^3.\label{eq:H_num}
\end{equation}
This value must be compared to the rate of gauge interactions, that
insure thermalization after reheating. Their scattering rate is
approximately given by $\Gamma_{gauge} \approx \alpha^2 T$, so that
the requirement $\Gamma_{gauge} > H$ leads to the following upper
limit on the reheat temperature and thus on the RHN mass:
\begin{equation} \label{eq:Tmax}
M\simeq  T < 3.4\times 10^5 \left(\frac{\sqrt{g_{*r}}\, T_r}{ 3.28 \mbox{ MeV}} \right)^{1/2}
\mbox{ GeV},
\end{equation}
for $\alpha=1/30$. The bound of Eq.~(\ref{eq:Tmax}) implies that in
the quintessential kination scenario thermal leptogenesis is viable
for masses which are typically lower than in the standard case. Of
course, the above constraint on $M$ can be relaxed when $T_r>>1$ MeV,
as the conventional picture of radiation domination is recovered for
$z_r \to 0$.  Moreover, in order to allow the conversion of the lepton
asymmetry to the baryon asymmetry, the electroweak sphaleron
interaction must be in thermal equilibrium before the electroweak
phase transition.  Taking the order of magnitude estimate of the
sphaleron interaction rate $\Gamma_{sp} \sim \alpha^4 T$
\cite{sprate}, the condition $\Gamma_{sp} > H$ reduces the above
constraint (\ref{eq:Tmax}) by an additional factor $\alpha$:

\beq
T < 10^4 \left(\frac{\sqrt{g_{*r}}\, T_r}{ 3.28 \mbox{ MeV}}
\right)^{1/2}
\mbox{ GeV} .
\eeq

\noindent Therefore, we get a sufficient window
for the sphaleron transition to thermalize above the electroweak
temperature $T\sim 100$ GeV even for $T_r$ = 1 MeV .

\section{Supersymmetric kination leptogenesis}
\label{sec:themodel}

In this paper we wish to discuss thermal leptogenesis in the Minimal
Supersymmetric extension of the Standard Model (MSSM) supplemented by
right--handed neutrino supermultiplets, i.e. the model described by
the following superpotential:

\beq
{\cal W}={\cal W}_{MSSM}+\frac{1}{2} N^c M N^c+y H_2 L N^c.
\label{eq:superpotential}
\eeq
This scenario has  been extensively studied in the literature
\cite{Plumacher,Giudice} in a conventional cosmological setup where
the energy density of the early Universe is dominated by radiation. We
wish now to discuss leptogenesis in the scenario of kination
domination introduced in the previous section.  Let us remark that
our results can also be applicable to the non-supersymmetric case
without any qualitative change.

In the MSSM, the decay rate of a RHN is given by $\Gamma_d = |y|^2
M/4\pi$ where $y$ is the neutrino Yukawa coupling. Introducing as
usual the effective neutrino mass scale given by
\begin{equation}
 \tilde{m} \equiv |y|^2 {\frac{\langle H_2 \rangle^2}{M}},
\end{equation}
\noindent and using Eq.~(\ref{eq:H_zr}), we can get the ratio
$K\equiv \Gamma_d/H (z=1)$:
\beq
K=\frac{63.78}{\sqrt{1+z_r^2}}\left (\frac{\tilde{m}}{0.05\; {\rm ev}}
\right).
\label{eq:K_zr}
\eeq
We remind that the standard cosmology is recovered in the above equation
when $z_r<<1$, and thus leptogenesis proceeds in the strong wash--out
regime, $K>>1$, for $\tilde{m}$ equal to the observed atmospheric
neutrino mass scale.  The situation changes for kination domination,
where $z_r >> 1$.  In particular, putting numbers in the above
equation one gets:
\begin{equation}
 K = 1.38 \times 10^{-6} \left(\frac{ 10^4 \mbox{ GeV}}{M} \right)
 \left( \frac{\tilde{m}} {0.05 \mbox{ eV}} \right)
 \left(\frac{ \sqrt{g_{*r}}\, T_r} {3.28 \mbox{ MeV}}\right) .
\label{eq:K_kination}
\end{equation}
\noindent
Thus, if kination dominates the energy density of the Universe until
nucleosynthesis, leptogenesis is in the super-weak wash-out regime,
$K<<<1$, even when $\tilde{m}\simeq$ 0.05 eV. On the other hand,
assuming $T_r>> 1$ MeV, higher values of $K$ can be attained, so that
a broad range of different scenarios are possible. However, as will be
shown in the discussion of Section \ref{sec:discussion}, an upper
bound $K\lsim 10$ exists if kination has to play any r\^ole for
leptogenesis.

The Boltzmann equations that drive leptogenesis involve the comoving
densities $Y_i(z=M/T)\equiv n_i/s(z)$, with $n_i$ the density of the
species $i$, $s$ the entropy density, and $i$=$N$ (heavy Majorana
neutrinos), $\tilde{N}$ (sneutrinos), $\tilde{N}^{\dagger}$
(anti--sneutrinos), $l$ (leptons), $\bar{l}$ (anti--leptons),
$\tilde{l}$ (sleptons) and $\tilde{l}^{\dagger}$ (antisleptons). The
additional degrees of freedom $h_2,h^{\dagger}_2$ (Higgs bosons) and
$\tilde{h}$ (higgsino) need not to be studied, since their densities
are constrained to be equal to $\tilde{l},\tilde{l}^{\dagger}$ and
$l$.  All other light degrees of freedom are assumed in thermal
equilibrium.  It is convenient to introduce the following adimensonal
densities:
\begin{eqnarray}
&&N(z)\equiv \frac{Y_N(z)}{Y^{eq}_N(z=0)}\,, \quad
\tilde{N}(z)\equiv
\frac{Y_{\tilde{N}}(z)}{Y^{eq}_{\tilde{N}}(z=0)}\,,\quad
\tilde{N^{\dagger}}(z)\equiv
\frac{Y_{\tilde{N^{\dagger}}}(z)}{Y^{eq}_{\tilde{N^{\dagger}}}(z=0)}\,, \nonumber\\
&&
\tilde{N}_\pm\equiv\tilde{N}(z)\pm\tilde{N}^{\dagger}(z)\,,\quad
L(z)\equiv \frac{Y_l(z)-Y_{\bar{l}}(z)}{Y^{eq}_l(z=0)}\,,\quad
\tilde{L}(z)\equiv \frac{Y_{\tilde{l}}(z)-Y_{\tilde{l}^{\dagger}}(z)}{Y^{eq}_{\tilde{l}}(z=0)}
\,.
\end{eqnarray}
The Boltzmann equations of the system can be simplified by noticing that MSSM
gaugino--mediated interactions of the type $l+l\rightarrow \tilde{l}
\tilde{l}$ are very fast and can be safely assumed to be in thermal
equilibrium. This automatically implies that $\tilde{N}_-$=0 and
$L$=$\tilde{L}$. Thus setting now
\begin{eqnarray}
\hat{N}(z)&\equiv& N(z)+\tilde{N}_+ \nonumber\\
\hat{L}(z)&\equiv& L(z)+\tilde{L}(z),
\end{eqnarray}
the Boltzmann equations driving the system are given by
\bea \frac{d \hat{N}}{dz}(z)&=&-K \sqrt{\frac{1+z_r^2}{z^2+z_r^2}} z^2
 (\hat{N}-\hat{N}_{eq})\left [\gamma_d(z)+2 \gamma_s(z) +
4 \gamma_t(z)\right] \label{eq:boltzman_susy_zr_n} \\
\frac{d \hat{L}}{dz}(z)&=&K
 \sqrt{\frac{1+z_r^2}{z^2+z_r^2}}
z^2\left [\gamma_d(z) \epsilon (\hat{N}-\hat{N}_{eq})-\frac{\gamma_d(z)
\hat{N}_{eq}\hat{L}}{4}-\frac{1}{2}\gamma_s(z) \hat{L}\hat{N}-\gamma_t(z)
\hat{L}\hat{N}_{eq}\right ]\label{eq:boltzman_susy_zr_l}.
 \eea
\noindent where the CP--violating parameter $\epsilon$ is defined as
in Ref.~\cite{Plumacher}, $\gamma_d(z)=K_1(z)/K_2(z)$ ($K_{1,2}$ are
Bessel functions of the first kind) and $\hat{N}_{eq}(z)=z^2
K_2(z)/2$.  Here, we have included the dominant lepton--number
violating scattering amplitudes $\gamma_{s,t}$ proportional to the top
Yukawa coupling $\lambda_t$ and driven by Higgs exchange, which are
given by:
\bea \gamma_{s,t}(z)&\equiv& \frac{1}{n_{eq} \Gamma_d}\frac{T}{64
\pi^4}\int ds\; \hat{\sigma}_{s,t}(s) \sqrt{s}
K_1(\frac{\sqrt{s}}{T})\\
\mbox{where}\quad
n_{eq}(z)&=&\frac{g}{2
\pi^2}{M^3 \over z} K_2(z)\;\;
\mbox{with}\;\; g=2 \;({\rm either \;\; Majorana}\;\; \nu\;\; {\rm
or}\;\;\; \tilde{\nu}+\tilde{\nu}^{\dagger}) \nonumber \\
\mbox{and}\quad
\hat{\sigma}_{s,t}(s)&\equiv& 3 \frac{\alpha_u}{4 \pi} f_{s,t}\left
(\frac{s}{M^2}\right ),\;\;\;\alpha_u=\frac{\lambda_t^2}{4 \pi}. \nonumber
\label{eq:gammas}
\eea
\noindent In the above equations the functions $f_{s,t}$ are given by:
\bea f_s(x)&\equiv& 3 \left
[f^{(0)}(x)+\frac{f^{(3)}(x)}{2}+f^{(5)(x)}+\frac{f^{(8)}(x)}{2}
+\frac{f_{22}(x)}{2} \right ]\nonumber \\ f_t(x)&\equiv& \frac{3}{2}\left
[f^{(1)(x)}+f^{(2)}(x)+f^{(4)}(x)+f^{(6)}(x)+f^{(7)}(x)+f^{(9)}(x)+f_{22}(x),
\right ].\label{eq:ff} \eea
\noindent where the functions $f^{(i)}$ can be read--off from
Eqs.~(C.20-C.30) and Eq.~(14) of Ref.~\cite{Plumacher}. In our
calculation, we include the thermal effect by taking $a_h\equiv
m_H(T)/M$ \cite{Giudice}, where $m_H(T)\simeq 0.4\; T$ is the thermal
mass for the Higgs/Higgsino particles. We remark that in our analysis
the difference between bosons and fermions is neglected, as we work in
the Boltzmann approximation.  The main effect of the quantity $a_h$ is
to regularize the infrared divergence in the t--channel, which shows
up in the logarithms in the functions $f^{(i)}$.  In particular, when
$a_h$ is neglected everywhere with the exception of the logarithms,
the functions $f_{t,s}$ take the simplified form:
\bea f_s&\equiv& 6 \;\frac{x-1}{x} \label{eq:fs_approx}\nonumber\\
f_t&\equiv&
6\left [-1+ \log \frac{x-1+a_h}{a_h}\right ].\label{eq:ft_approx}
\label{eq:f_s_t_simple}
\eea
However, in our numerical calculations, we kept $a_h$ everywhere
(although the above simplified form leads to very similar
conclusions). Note that scattering effects may be safely included at
the tree--level, and at this order of perturbation theory they are
CP--conserving processes.

\begin{figure}
\begin{center}
\hspace{-1cm}
\includegraphics*[height=9cm, bb=31 199 514 632]{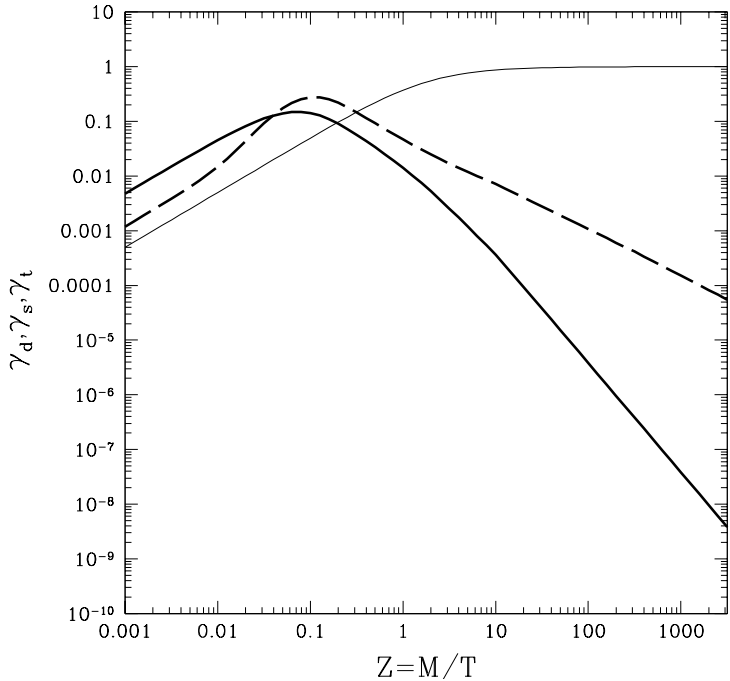}
\end{center}
\caption{Decay and scattering rates $\gamma_d$ (thin solid line),
$\gamma_s$ (thick solid line), $\gamma_t$ (thick dashed lines) used in
the Boltzmann equations
(\protect\ref{eq:boltzman_susy_zr_n},\protect\ref{eq:boltzman_susy_zr_l}),
as a function of $z$.
\label{fig:rates_susy}}
\end{figure}

The effect of thermal corrections on the other masses, as well as on
the CP--violating parameter $\epsilon$, has been considered in the
literature \cite{Giudice}, and can be important, especially at high
temperatures.  However, in our case
the bulk of leptogenesis takes place at temperatures low enough to
suppress any sizable effect on the final result, so in the following
we will neglect thermal effects, with the exception of the
introduction of the coefficient $a_h(T)$.  The functions $\gamma_d$ and
$f_{s,t}$ are plotted in Fig.~\ref{fig:rates_susy} as a function of
$z$. Fig.~\ref{fig:rates_susy} shows that the
effect of scattering is particularly important at higher temperatures,
where the amplitudes $\gamma_{s,t}$ are larger than $\gamma_d$. On the
other hand, at lower temperatures the effect of scattering turns out
to be negligible.

\section{Properties of leptogenesis: kination vs.\ radiation}
\label{sec:discussion}

In this section we discuss the solutions of the Boltzmann
Eqs.~(\ref{eq:boltzman_susy_zr_n},\ref{eq:boltzman_susy_zr_l}) for
some illustrative examples.  We mainly concentrate here on the case of
a vanishing initial RHN density, $\hat{N}(0)=0$, and give some
comments also for the case $N(0)=1$ (initial thermal distribution for
the RHN).  Besides solving them numerically for the more general
cases, it is useful to work--out semi--analytic solutions of
Eqs.~(\ref{eq:boltzman_susy_zr_n},\ref{eq:boltzman_susy_zr_l}) in the
limits $K<<1$ and $K>>1$ in the two cases of radiation domination
($z_r<<1$) and kination domination ($z_r>>1$). \\

\begin{figure}
\begin{center}
\hspace{-1cm}
\includegraphics*[height=7cm, bb=31 199 514 632]{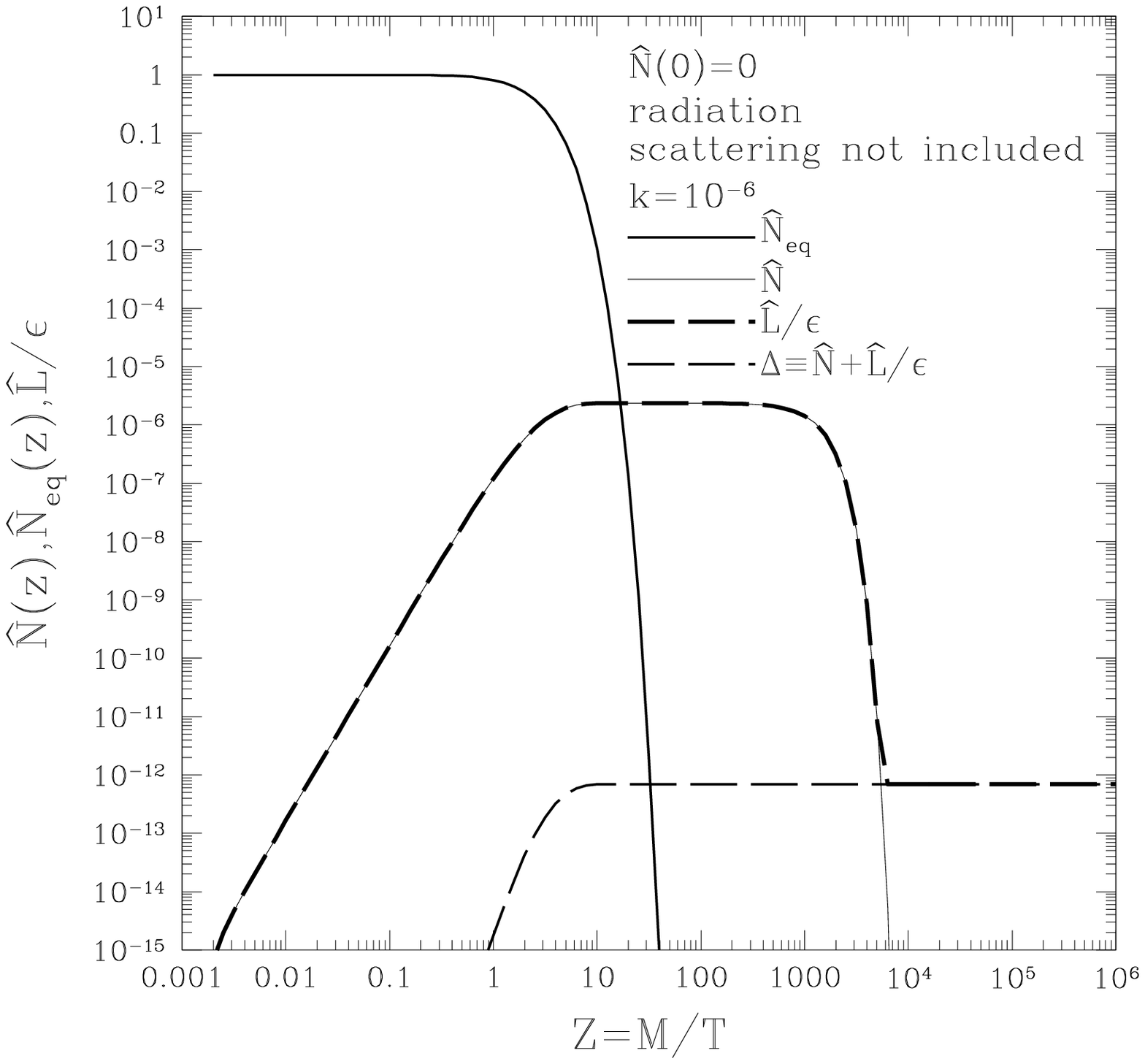}
\includegraphics*[height=7cm, bb=31 199 514 632]{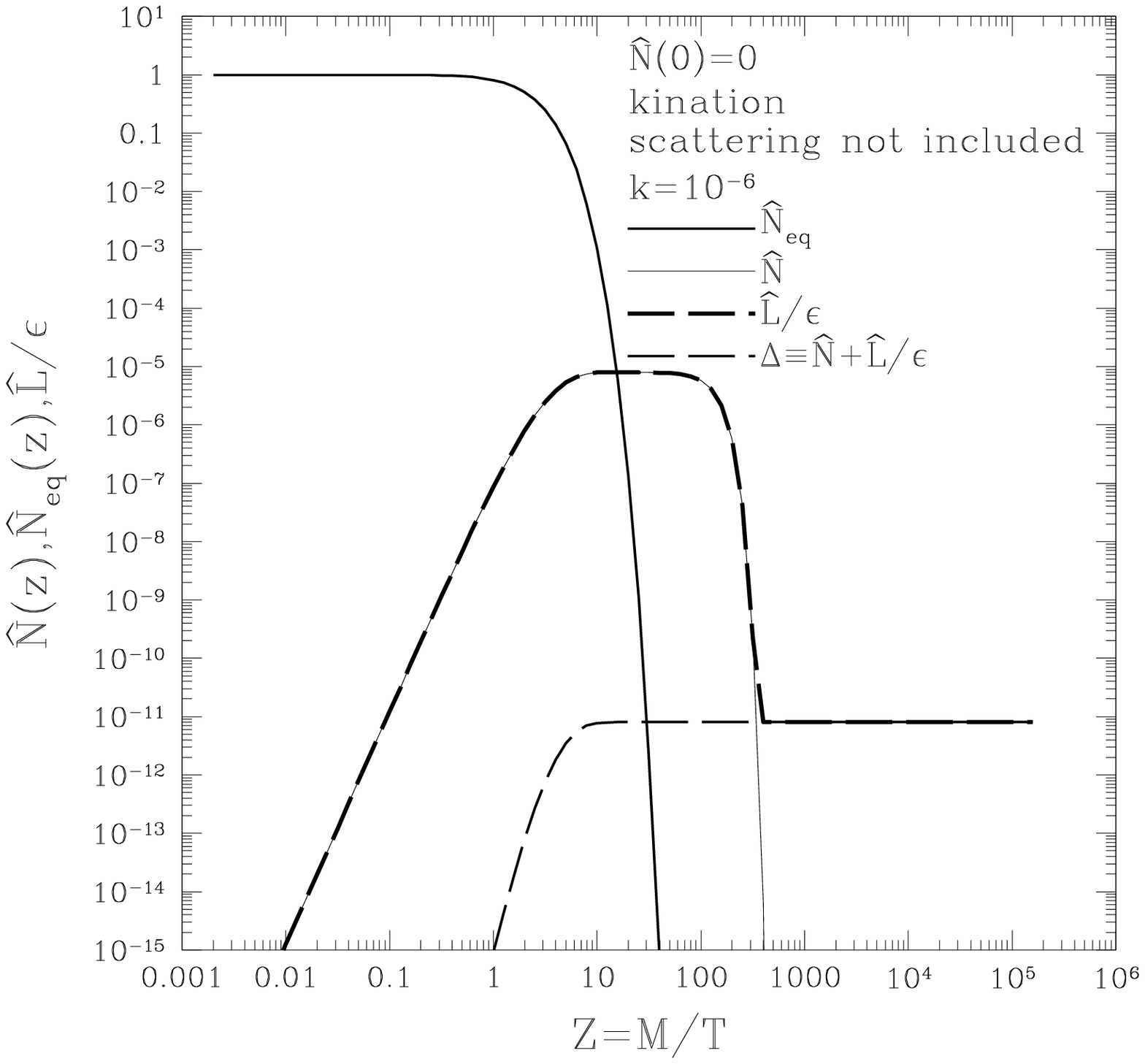}
\end{center}
\caption{Evolution as a function of $z\equiv M/T$ of the quantities
$\hat{N}$, $\hat{L}/\epsilon$, $\Delta\equiv \hat{N}+\hat{L}/\epsilon$, for the case
$K=10^{-6}$, and neglecting scattering. Left panel: energy density dominated by the radiation
field; right panel: energy density dominated by kination.
\label{fig:example_no_scattering_k_1d-7}
}
\end{figure}

\begin{figure}
\begin{center}
\hspace{-1cm}
\includegraphics*[height=7cm, bb=31 199 514 632]{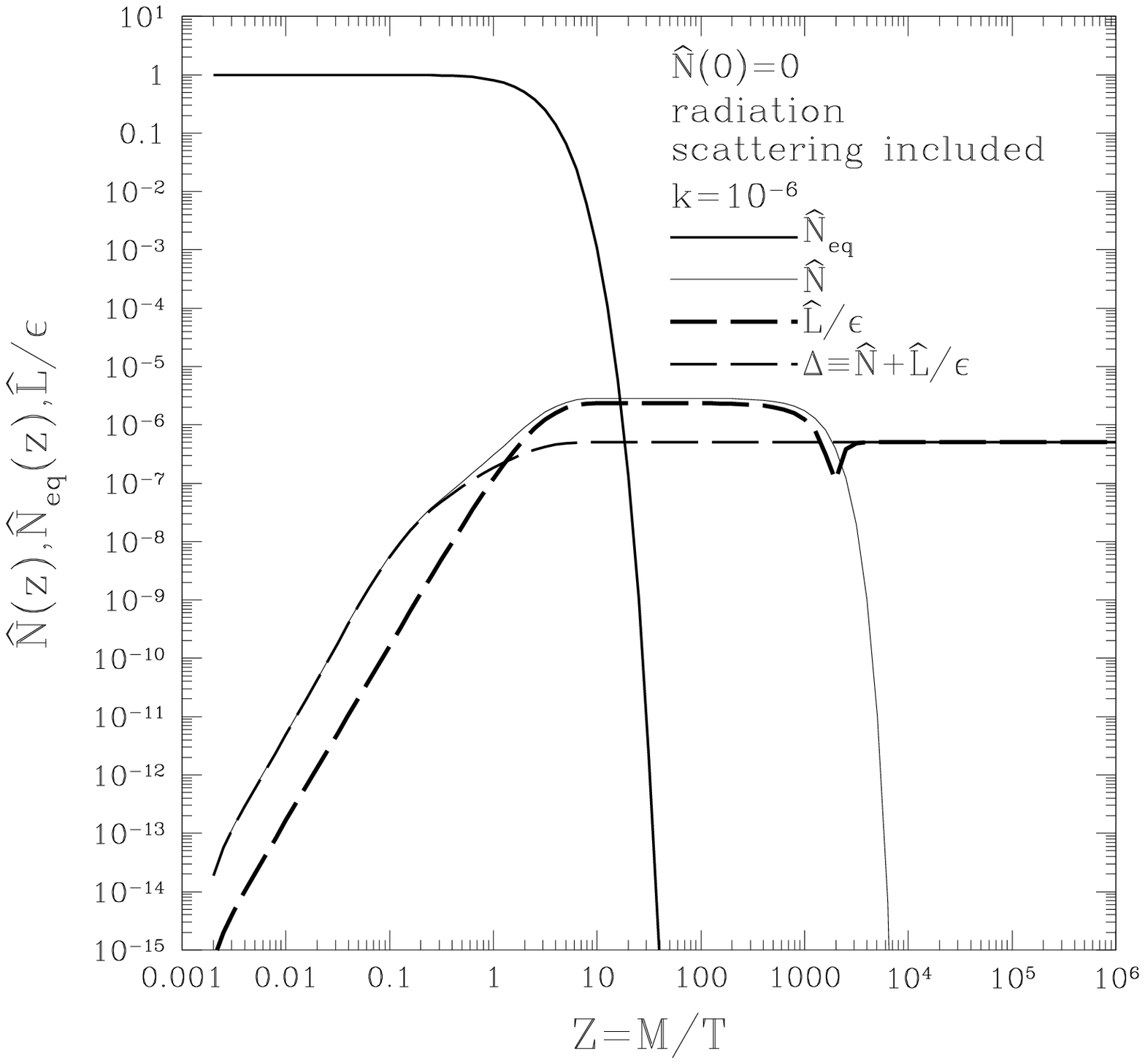}
\includegraphics*[height=7cm, bb=31 199 514 632]{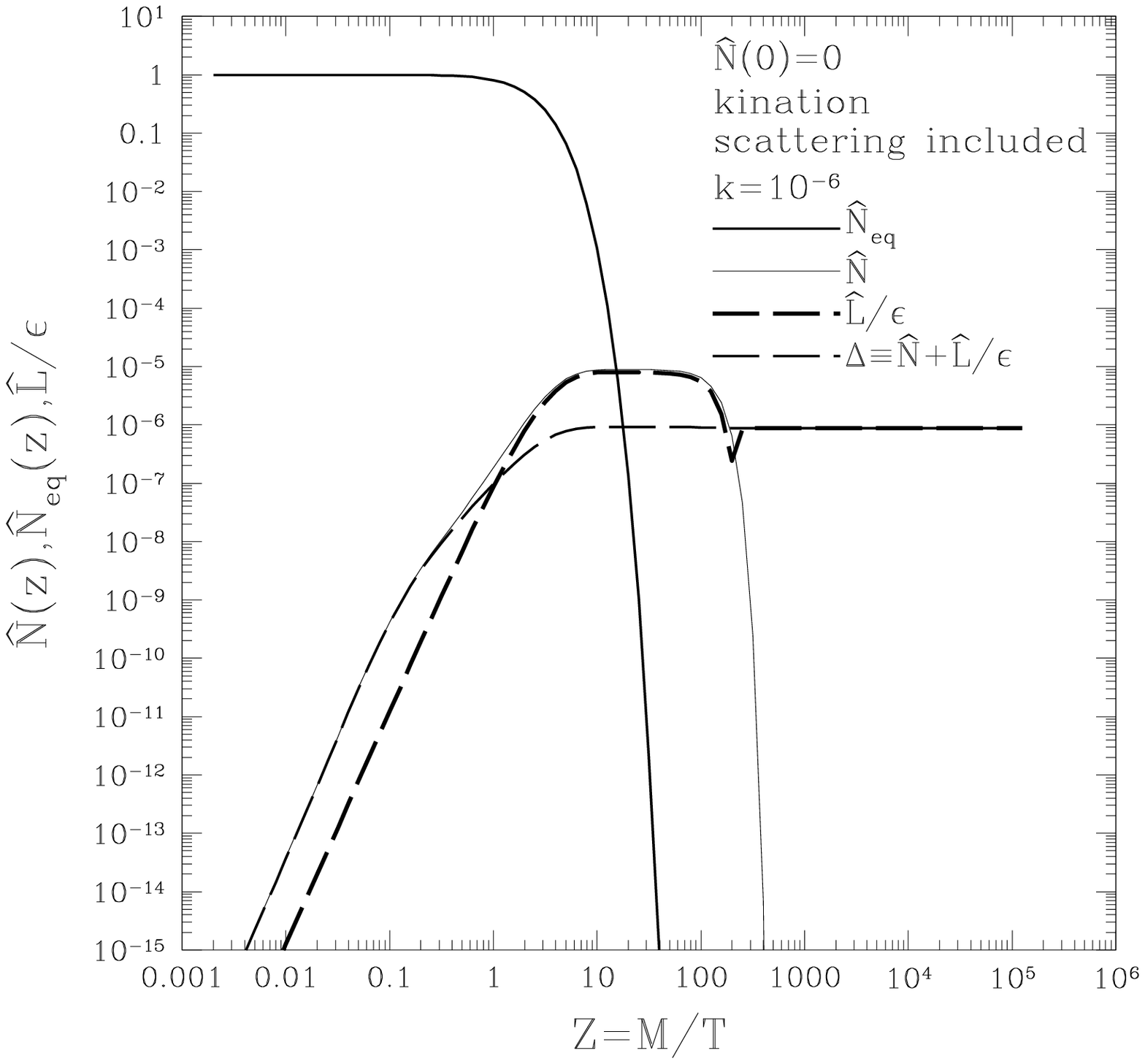}
\end{center}
\caption{The same as in
Fig.~\protect\ref{fig:example_no_scattering_k_1d-7} with the inclusion
of the scattering effect.
\label{fig:example_scattering_k_1d-7}
}
\end{figure}

\noindent
{\bf (i) Super--weak wash--out regime.}\\ In the case $K<<1$,
leptogenesis takes place early, when $z\lsim 1$ (see discussion of
Figs. \ref{fig:example_no_scattering_k_1d-7} and
\ref{fig:example_scattering_k_1d-7}). As a consequence of this, in
Eq.(\ref{eq:boltzman_susy_zr_n}) one can neglect $\hat{N}\ll
\hat{N}_{eq}$, so that it decouples from
Eq.(\ref{eq:boltzman_susy_zr_l}). Setting:

\beq \Delta\equiv
\hat{N}+\frac{\hat{L}}{\epsilon},
\label{eq:Delta}
\eeq

\noindent one gets for the efficiency $\eta$ the solution:

\begin{eqnarray}
\eta&\equiv & \frac{\hat{L}(\infty)}{\epsilon}=\Delta(\infty)\simeq
K\int_0^{\infty} z^{n} \hat{N}_{eq}(z)(\gamma_d + 2 \gamma_s+4
\gamma_t) \; dz\simeq
K \int_0^{\infty} z^{n+2} K_2(z)(\gamma_s+2
\gamma_t) \nonumber\\
&=&K \hat{I}_n \quad
\mbox{with} \quad
\hat{I}_1\simeq 0.504,\;\mbox{and}\;\; \hat{I}_2\simeq 0.921,
\label{eq:delta_analytic}
\end{eqnarray}
\noindent where $n=1$ and 2 for radiation and kination domination,
respectively.
In the second integral of Eq.~(\ref{eq:delta_analytic}) $\gamma_d$ has
been dropped, since the integral takes its main contribution at
$z\lsim 1$ where scattering amplitudes dominate.  So, for $K<<1$,
scattering turns out to be essential in determining the efficiency
$\eta$.  Eq.~(\ref{eq:delta_analytic}) is consistent with the result
of Ref.~\cite{buch}, which in particular showed how $\eta$ is
proportional to $K^2$ or $K$, depending on whether scattering terms
are neglected or included, respectively.

To see more in detail the behavior of the numerical solutions of
Eqs.~(\ref{eq:boltzman_susy_zr_n},\ref{eq:boltzman_susy_zr_l}) in the
two cases of kination-- and radiation--domination, we show them for
$K=10^{-6}$ in Figs.~\ref{fig:example_no_scattering_k_1d-7} and
\ref{fig:example_scattering_k_1d-7}. In particular, in the former the
scattering effect is neglected, while in the latter it is included. In
both figures the evolution of the different quantities $\hat{N}$,
$\hat{L}/\epsilon$ and $\Delta$ is shown as a function of $z$, and the
left hand panel shows the result obtained by assuming $z_r<<1\; (n=1)$
in the Boltzmann equations (radiation domination), while the right
hand panel is calculated for $z_r>>1\; (n=2)$ (kination dominated).

Some general features, valid for both kination and radiation, can be
realized by comparing these figures. In both cases, at higher
temperatures an initial population of RHNs and an early lepton
asymmetry are built up. This process is active as long as the initial
light states have enough kinetic energy to create RHNs, i.e., as long
as $z=M/T<1$. By the time when $z\simeq 1$ and this process stops,
RHNs have neither thermalized nor decayed due to their very weak
couplings, so both the RHN density and the lepton asymmetry are frozen
as can be seen from their evolution in
Figs.~\ref{fig:example_no_scattering_k_1d-7} and
\ref{fig:example_scattering_k_1d-7} which show a plateau until much
later ($z\gg 1$).

In Fig.~\ref{fig:example_no_scattering_k_1d-7} where the scattering
processes are not included in the calculation, one can see that the
asymmetry $\hat{L}/\epsilon$ tracks closely $\hat{N}$ until the decay
of RHNs. Moreover, when RHNs decay, they produce a lepton asymmetry
that cancels the one created earlier due to inverse decays by several
orders of magnitude. It is worth noticing here that inverse--decay
rates are proportional to the densities of initial states, so the
inverse--decay rate for a particle whose reaction is faster than its
anti--particle (due to CP violation) is suppressed by the higher
depletion in the corresponding population. This implies that the
effective amount of CP violation in the early inverse--decay processes
is slightly smaller than $\epsilon$. On the other hand, when the RHNs
decay out--of--equilibrium at later times the amount of CP violation
is exactly given by $\epsilon$. So the lepton asymmetry created later
has opposite sign and slightly overshoots the earlier one (actually,
the wiggle that is visible in the evolution of $\hat{L}/\epsilon$
signals a sign change, since it is plotted in absolute value).  It is
this mismatch that explains why in this plot the asymmetry created at
early times by inverse decays is not exactly canceled by RHN decays at
later times.  Anyway, the final value of the asymmetry is given by a
very strong cancellation, and is essentially produced by a
second--order back--reaction effect.  In particular, it can be seen
that, when scattering is neglected, $\hat{L}/\epsilon\propto K^2$
\cite{buch}, which implies a strong suppression for $K<<1$.

The inclusion of scattering processes in the solution of the Boltzmann
equations changes the property described above. First of all, due to
the higher overall interaction rate, the RHNs are more populated in
the first place, and this enhances the final asymmetry. However, the
main effect which is active now is due to the presence of $s$--channel
scatterings of the type $Q+U\rightarrow N+L$. In fact, this last type
of interactions, which is dominant for $z\lsim 0.2$ (see
Fig.~\ref{fig:rates_susy}) is (approximately) a CP--conserving
process. This means that it populates $\hat{N}$ without affecting
$\hat{L}/\epsilon$ (since, for instance, $\Gamma(Q+U\rightarrow N
+L)=\Gamma(\bar{Q}+\bar{U}\rightarrow N +\bar{L}))$. As a consequence
of this, in both panels of Fig.~\ref{fig:example_scattering_k_1d-7}
now the asymmetry $\hat{L}/\epsilon$ no longer tracks $\hat{N}$ (this
may be appreciated in Fig.~\ref{fig:example_scattering_k_1d-7} at
early times, when $z\lsim 0.2$, and s--channel scattering processes
are dominant).  When these RHNs decay at later times, they produce a
lepton asymmetry that is not canceled by an earlier, specular one,
left over by their earlier production. This explains why in
Fig.~\ref{fig:example_scattering_k_1d-7} almost all the asymmetry
produced for $z\lsim 0.2$ survives until later times, while that
created later is washed out. As a consequence of this, including
scattering processes the efficiency is much higher and the
characteristic drop in the asymmetry when RHNs decay is much less
pronounced.

By comparing the two panels of
Fig.~\ref{fig:example_scattering_k_1d-7}, one can see the effect of
different cosmological models on the evolution of the lepton
asymmetry. The final values of the efficiency are quite similar in the
two cases. As estimated in Eq.~(\ref{eq:delta_analytic}), the
difference between kination and radiation in the final asymmetry is
expected to be about a factor of 2 for fixed $K$. Moreover, the
plateau in the evolution of the asymmetry corresponds to a shorter
interval in $z$ in the case of kination compared to radiation: this is
due to the fact that, in the time interval given by the RHN lifetime,
the Universe is decelerating faster for kination, implying less
expansion and cooling, so that the corresponding variation of the $z$
parameter is smaller. \\

\noindent{\bf (ii) Strong wash--out regime}:\\ For $K>>1$, the
semi--analytic solutions of the Boltzmann equations
(\ref{eq:boltzman_susy_zr_n},\ref{eq:boltzman_susy_zr_l}) can be
calculated using the saddle point technique. In this case the bulk of
the lepton asymmetry is produced at the the decoupling temperature
$z_f$, determined by the approximate relation:

\bea z_f^{n+\frac{3}{2}} e^{-z_f}=\frac{2^{\frac{7}{2}}}{K \pi^{1/2}},
\eea (where again $n=1$ for radiation and $n=2$ for kination) which
can be approximated by the logarithmic fits:
\begin{eqnarray}
&&z_f\simeq a_n+b_n \ln(K),\nonumber\\
&&a_1=1.46;\;\;\;b_1=1.40,\nonumber\\
&&a_2=4.66;\;\;\;b_2=1.41.
\label{eq:zf_fit_susy}
\end{eqnarray}
By comparing Eqs.~(\ref{eq:K_zr}) and (\ref{eq:zf_fit_susy}) it is
possible to set an upper bound on $K$, as anticipated in the previous
section. In fact, by fixing $K$ in both equations and requiring that
$z_f<z_r$ (i.e., that decoupling happens in the regime of kination
domination), one gets the inequality:
\beq
z_f\simeq a_2+b_2 \ln(K)<\sqrt{\left
  (\frac{63.78}{K}\frac{\tilde{m}}{0.05\;{\rm eV}} \right
  )^2-1}=z_r.
\label{eq:inequality}
\eeq
This numerically implies, for instance, $K\lsim$ 7.6 for
$\tilde{m}=0.05$ eV, and $K\lsim$ 5.7 for $\tilde{m}=0.01$ eV. The
inequality (\ref{eq:inequality}) has no solution for $K>1$ when
$\tilde{m}\lsim 0.004$ eV.

For $K>>1$, the final value of the efficiency is given by the
approximate expression:
\begin{eqnarray}
&&\eta
=\frac{\hat{L}(\infty)}{\epsilon}=\hat{N}_{eq}(z_f)F_{wash-out}(n,z_f),
\label{eq:l_approx_large_k} \\
&&~~\mbox{where} \quad
\hat{N}_{eq}(z_f)=\frac{4}{K z_f^n}, \label{eq:n_eq_zf_susy} \\
&&~~~~\mbox{and} \quad
 F_{wash-out}(n,z_f)\simeq
\sqrt{\frac{2\pi}{1-\frac{n}{z_f}}}\, \exp\left [
-\left(1+\frac{\frac{3}{2}+n}{z_f} \right) \right ].
\label{eq:f_washout}
\end{eqnarray}

\begin{figure}
\begin{center}
\hspace{-1cm}
\includegraphics*[height=7cm, bb=31 199 514 632]{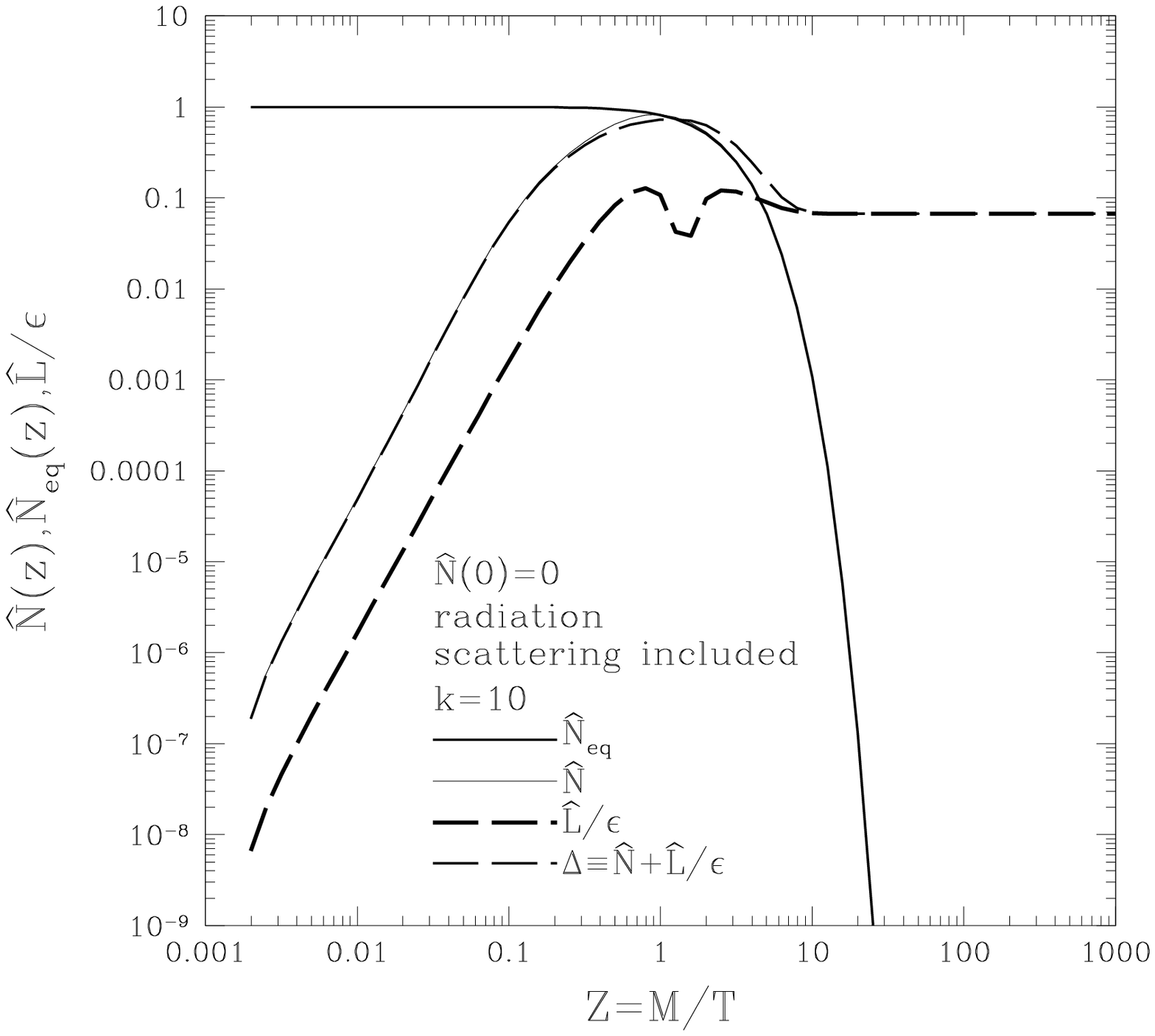}
\includegraphics*[height=7cm, bb=31 199 514 632]{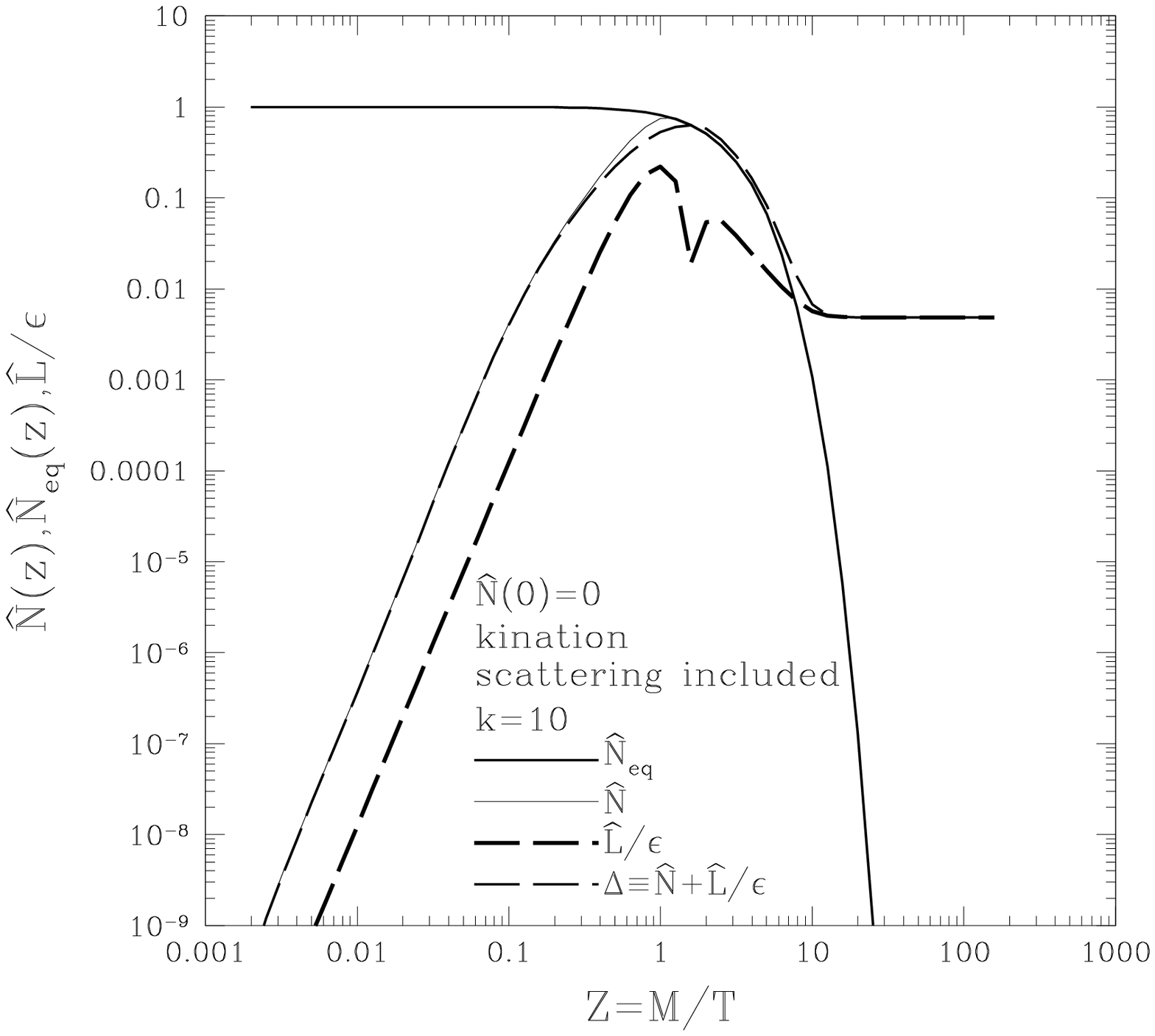}
\end{center}
\caption{The same as in
Fig.~\protect\ref{fig:example_scattering_k_1d-7}, for the case
$K=10$.\label{fig:example_k_10} }
\end{figure}

An example of the numerical solutions of
Eqs.~(\ref{eq:boltzman_susy_zr_n},\ref{eq:boltzman_susy_zr_l}) in the
strong wash--out regime is shown in Fig.~\ref{fig:example_k_10} for
$K=10$, in the case of kination domination (left panel) and radiation
domination (right panel).  As opposed to
Fig.~\ref{fig:example_scattering_k_1d-7}, now RHNs quickly thermalize,
and produce a lepton asymmetry when they decouple at $z\simeq
z_f>>1$. In Fig.~\ref{fig:example_k_10} the scattering effect is
included, but does not affect the final asymmetry, since the latter
may develop only at late times when the scattering processes are
negligible.

By comparing the two panels one can see that, as implied by
Eq.~(\ref{eq:zf_fit_susy}), RHNs stay in thermal equilibrium longer
for kination than for radiation. This is explained by the fact that
the two models have the same Hubble--expansion rate at $z=1$ (when $K$
is defined) but decoupling happens at $z>1$, when the model dominated
by kination is decelerating faster and has a slower expansion compared
to the case of radiation domination.  Moreover, the RHN equilibrium
density for $z>1$ is further suppressed for kination compared to
radiation [see Eq.~(\ref{eq:n_eq_zf_susy})]. This implies a sizeable
difference in the final lepton asymmetry between kination and
radiation, in contrast to what happens in the case $K<<1$ for which the
difference is less pronounced.  Note also that, since the RHNs
thermalize at $z\simeq$ 1, erasing any dependence of the final
asymmetry on earlier boundary conditions, the calculated efficiency
would be the same assuming $\hat{N}(0)$=1.

\begin{figure}
\begin{center}
\hspace{-1cm}
\includegraphics*[height=9cm, bb=31 199 514 632]{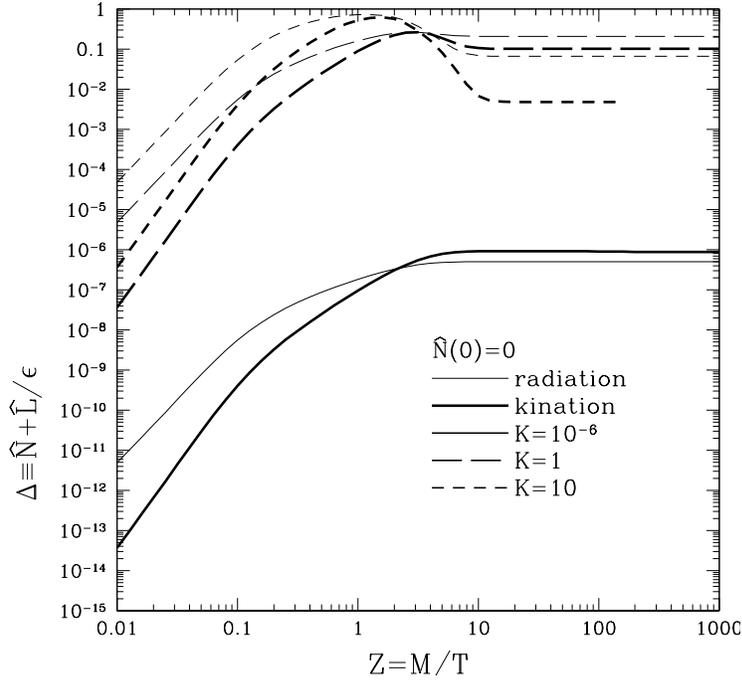}
\end{center}
\caption{Evolution of $\Delta\simeq \hat{N}+\hat{L}/\epsilon$ as a function of
  $z$, with the boundary condition $\hat{N}(0)=0$. Thin lines=radiation, thick
  lines=kination. Solid line, long dashes and short dashes for
  $K=10^{-6}$, $K=1$ and $K=10$, respectively.
\label{fig:scattering}}
\end{figure}

\begin{figure}
\begin{center}
\hspace{-1cm}
\includegraphics*[height=9cm, bb=31 199 514 632]{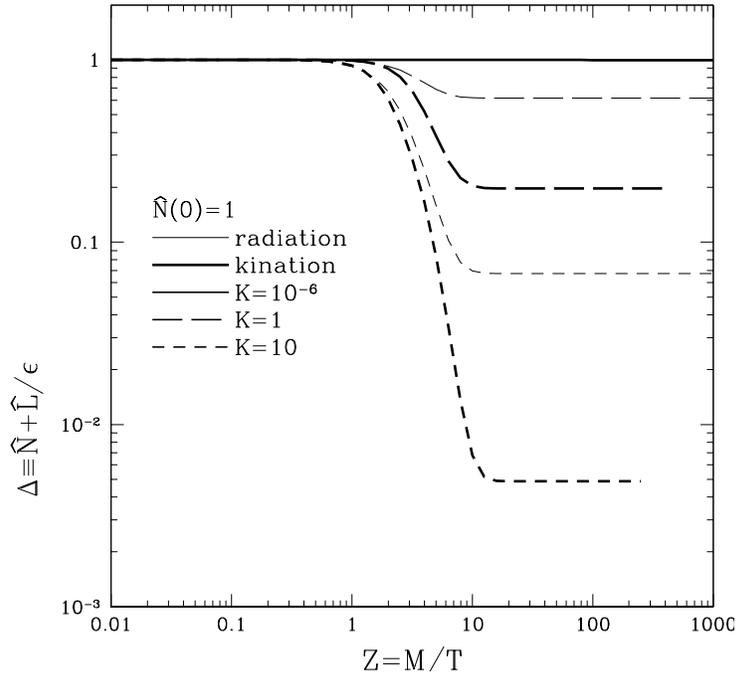}
\end{center}
\caption{The same as in Fig.~\protect\ref{fig:scattering}, for
$\hat{N}(0)=1$. \label{fig:scattering_eq}}
\end{figure}

In order to summarize the temperature--dependence of the solutions of
the Boltzmann equations
(\ref{eq:boltzman_susy_zr_n},\ref{eq:boltzman_susy_zr_l}), we present
in Fig.~\ref{fig:scattering} the evolution of $\Delta$ as a function
of $z$ for $K=10^{-6}$ (solid line), $1$ (long dashes), $10$ (long
dashes), in the case of radiation domination (thin lines) and kination
domination (thick lines), and when the boundary condition
$\tilde{N}(0)=0$ is adopted.  The same plot for the case
$\hat{N}(0)=1$ is shown in Fig.~\ref{fig:scattering_eq}.  In the
latter case, for $K<<1$ one has $\Delta$=1 during the whole range of
$z$, and both for radiation and kination. This corresponds to the
ideal case when the initial equilibrium density of RHNs decay out of
equilibrium with efficiency 1. For $K\gsim 1$ and as long as the RHNs
reach thermalization, the efficiency is the same as for the case with
$\hat{N}(0)=0$ already discussed. In particular, as already mentioned,
in this case the efficiency for kination is lower compared to
radiation because of the later decoupling for the RHNs.

\medskip

\begin{figure}
\begin{center}
\includegraphics*[height=8.7cm, bb=31 199 514 632]{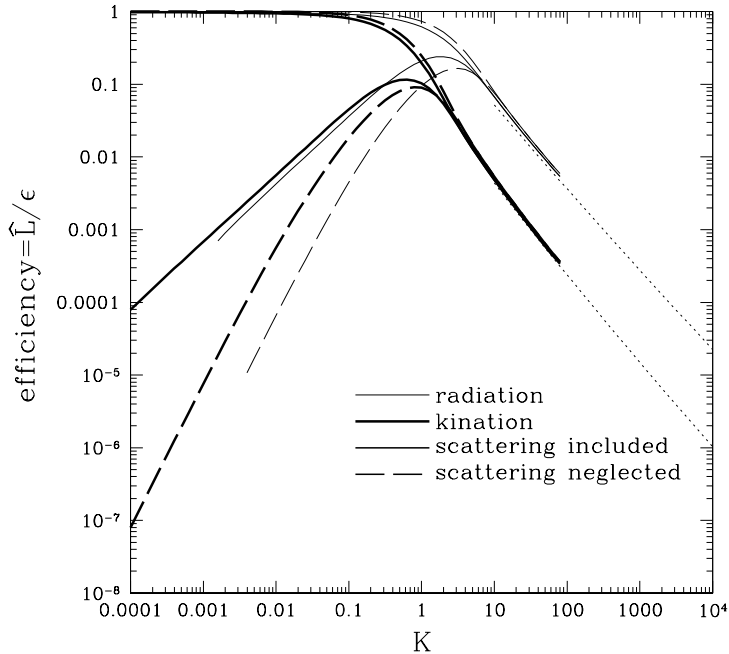}
\end{center}
\caption{The efficiency $\eta=\hat{L}(z=\infty)/\epsilon$ as a
function of $K$. Curves that saturate to 1 at $K\ll 1$ have a thermal
initial distribution ($\hat{N}(0)=1$) for RHNs, while those that
vanish when $k\ll 1$ have $\hat{N}(0)=0$.  Thick lines: kination;
thin lines: radiation; solid lines: scattering included; long
dashes: scattering neglected. Dots: approximations given for $k\gg
1$ by Eqs.~(\protect\ref{eq:zf_fit_susy},
\protect\ref{eq:l_approx_large_k}, \protect\ref{eq:n_eq_zf_susy},
\protect\ref{eq:f_washout}). \label{fig:efficiency_susy}}
\end{figure}

\noindent{\bf (iii) The efficiency in terms of $K$ or $(z_r,
\tilde{m})$}:

In Fig.~\ref{fig:efficiency_susy} the efficiency $\hat{L}/\epsilon$ is
plotted as a function of $K$. From this figure it is possible to
summarize the various properties discussed above, comparing
the cases of kination and radiation domination:

\begin{itemize}

\item curves for $\hat{N}(0)=0$ and $\hat{N}(0)=1$ differ at $K<1$
(the latter saturating to 1) but coincide for $K>1$;

\item the effect of scattering is important for $K<1$, implying in
  particular a large increase of the efficiency at very low $K$, but is
  negligible at $K>1$;

\item for $K>1$ the efficiency for kination is about one order of
  magnitude smaller than for radiation [see Eqs.~(25, 28)]; on the
  other hand the efficiencies are comparable in the two cases for
  $K\lsim 1$ [see Eq.~(23)].

\end{itemize}

\begin{figure}
\begin{center}
\hspace{-1cm}
\includegraphics*[height=9cm, bb=34 189 517 635]{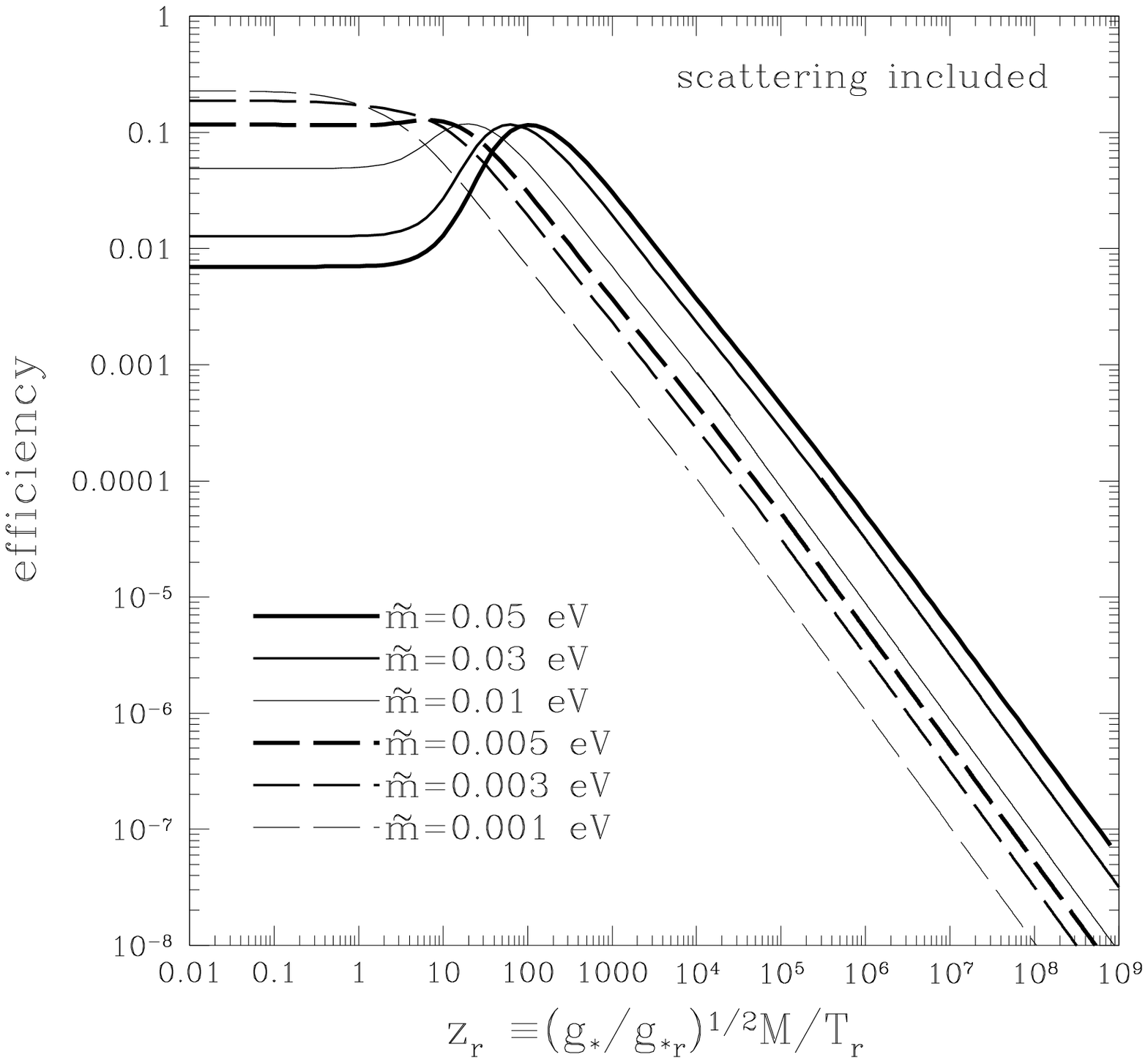}
\end{center}
\caption{The efficiency as a function of $z_r \equiv
  \sqrt{g_*/g_{*r}} M/T_r$, for $N(0)=0$ and for several values of
  $\tilde{m}$.  A smooth transition between radiation domination (the
  plateau at $z_r\lsim 1$) and kination-domination
  ($z_r \gsim 100$) is clearly visible.
\label{fig:efficiency_zr_smooth}}
\end{figure}

We conclude this section by discussing the phenomenology of
leptogenesis as a function of $T_r$, defined as the temperature for
which the energy density of radiation and kination are the same.  In
Fig.~\ref{fig:efficiency_zr_smooth} the efficiency is plotted as a
function of the adimensional parameter $z_r\equiv
\sqrt{\frac{g_*}{g_{*r}}}M/T_r$, for some representative values of the
neutrino mass scale $\tilde{m}$.  In this figure it is possible to see
a smooth transition from radiation domination ($z_r<<1$) to kination
domination ($z_r>>1$), and the following characteristic behaviors:

\medskip

\begin{itemize}
\item super--weak wash--out regime with $z_r >> 100$. Assuming
$T_r\gsim$ 1 MeV and taking into account the corresponding constraint
$M\lsim 10^5$ [see Eq.~(\ref{eq:Tmax})] one gets the upper bound
$z_r\lsim 4.5 \times 10^8$. Assuming for instance $\tilde{m}=0.05$ eV,
this value of $z_r$ corresponds in Fig.~\ref{fig:efficiency_zr_smooth}
to $K\lsim\hat{L}/\epsilon\simeq 10^{-7}$.  This situation has to be
compared to the standard picture in which the energy density is
dominated by radiation, corresponding, for the same curve, to the
plateau at $z_r<<1$, for which $K\simeq 64$ and
$\hat{L}/\epsilon\simeq$ 0.07. So, if kination dominates the energy
density of the Universe until nucleosynthesis, the corresponding
efficiency for leptogenesis is suppressed by almost six orders of
magnitude compared to the standard cosmology with radiation
domination;

\item increased efficiency with $ 1 \lsim z_r \lsim 100$. However,
 from the same figure it is possible to see that very different
 scenarios are possible by assuming $T_r>>1$ MeV.  For instance, when
 $\tilde{m}\gsim$ 0.01 eV the efficiency in the case of kination can
 be even larger compared to radiation for $1\lsim z_r \lsim 100$.
 This can be explained by the fact that, for this range of
 $\tilde{m}$, the radiation dominated situation corresponds to a
 regime of strong washout, $K>10$, while for kination it is possible
 to have $K\simeq$ 0.1--1, i.e. right in the interval that corresponds
 to a maximal efficiency. For instance, in
 Fig.~\ref{fig:efficiency_zr_smooth} the maximum at $z_r\simeq$100 for
 the curve with $\tilde{m}$=0.05 eV corresponds to $K\simeq$ 0.63,
 which in Fig.~\ref{fig:efficiency_susy} is the value of $K$ where the
 efficiency is maximal for kination domination. On the other hand,
 this mechanism gets much more efficient and $\hat{L}/\epsilon$ is
 strongly suppressed for $z_r>>10^3$. Note that the possibility to
 increase the efficiency in the case of kination domination compared
 to radiation is easier for somewhat larger values of $\tilde{m}$. In
 fact, if one takes $\tilde{m}<$ 0.05 eV, the value of $K$ in the
 radiation dominated case becomes smaller. According to
 Fig.~\ref{fig:efficiency_susy}, when $K\simeq$ 1 the efficiency for
 radiation is maximal, and this corresponds to $\tilde{m}\simeq
 8\times 10^{-4}$ eV. So, for $\tilde{m}< 8\times 10^{-4}$ eV adding a
 period of kination always implies a lower efficiency.

\end{itemize}

\section{Conclusions}
\label{sec:conclusions}

We have discussed leptogenesis in the context of a
cosmological model where the energy density of the early Universe is
dominated by the kinetic term of a quintessence field during some
epoch of its evolution. This assumption may lead to very different
conclusions compared to the case of the standard cosmology, where the
energy density of the Universe is dominated by radiation.

We have adopted as a free parameter the temperature $T_r$ above which
kination dominates over radiation, and shown that when $T_r$ is set to
its minimal value required by nucleosynthesis, $T_r\simeq$ 1 MeV,
gauge interactions can thermalize only at a temperature $T\lsim
10^{5}$ GeV, so that the RHN mass $M\simeq T$ needs to be relatively
light. This constraint is relaxed for higher values of $T_r$.
Moreover, irrespective of $T_r$, we always find a sufficient window
above the electroweak temperature $T\sim 100$ GeV for the sphaleron
transition to thermalize and thus allow conversion of the lepton
asymmetry to the observed baryon asymmetry.

When the RHN Yukawa coupling is fixed to get the observed neutrino
mass scale $\simeq$ 0.05 eV, in standard cosmology leptogenesis
proceeds in the strong wash--out regime, $K>>1$. On the contrary, in
kination leptogenesis any situation between the strong and the
super--weak wash--out regime are equally viable by varying
$T_r$. However, the effect of kination turns out to be either
negligible or absent for models for which $K\gsim 10$ since the RHNs
decay when radiation domination has already settled.

The weak wash--out regime is attained when $z_r\simeq M/T_r >>$ 100
for which the final efficiency for leptogenesis is approximately
given by $\eta\simeq K\simeq (64/z_r) (\tilde{m}/{\rm 0.05 eV})$,
and can be several orders of magnitude smaller than that in the case
of radiation domination for fixed $\tilde{m}$.  In this case we have
stressed the importance of $s$--channel scatterings driven by the
top Yukawa coupling, which are the dominant process to create RHNs
at high temperatures, and so enhance the leptogenesis efficiency by
orders of magnitude in models where a vanishing initial RHN density
is assumed. We remark that the condition $\eta \gsim 5\times
10^{-8}$ for a successful leptogenesis $Y_{\hat{L}} =4\times
10^{-3}\epsilon\eta \approx 10^{-10}$ requires the upper bound on
the RHN mass $M \lsim 2.8 \times 10^5 {\rm GeV} (T_r/{\rm MeV})$,
which is comparable to the thermal leptogenesis condition Eq.
(\ref{eq:Tmax}) for $T_r \sim 1$ MeV.
On the other hand, when $1\lsim z_r\lsim$ 100 kination stops to
dominate at a time which is not much later than when leptogenesis
takes place: in this case, if $\tilde{m}\gsim$ 0.01 eV, leptogenesis
proceeds with $0.1\lsim K\lsim 1$, in a regime where the efficiency is
even better than the for the case of radiation domination.

We conclude that a wide range of possibilities opens up for
leptogenesis if a period of kination--domination is assumed in the
early Universe. The ensuing phenomenology can be parametrized as a
function of only two parameters, $z_r\equiv
\sqrt{\frac{g_*}{g_{*r}}}M/T_r$ and $\tilde{m}$. \\

{\bf Acknowledgment}:
We thank D. Chung for useful discussions.

\end{document}